\documentclass[
amssymb,aps,
prd,nofootinbib,floatfix,11pt,
]{revtex4}%
\pdfoutput=1

\usepackage{amsmath,amssymb,epsf}
\usepackage{graphicx,color}

\usepackage{hyperref}

\usepackage{amssymb}
\usepackage{amsmath}
\usepackage{mathrsfs}
\usepackage{graphics}
\usepackage{epsfig}
\usepackage{graphicx}                   
\usepackage{subfigure}
\usepackage{bm}                              
\usepackage{nicefrac}
\usepackage{soul}

\usepackage{float}
\usepackage{placeins}
\usepackage{caption}
\usepackage{dcolumn}              

\definecolor{cblue}{RGB}{100,5,255}
\definecolor{cred}{RGB}{255,50,40} 
\definecolor{cgreen}{RGB}{1,100,0} 

\def\lsim{\mathrel{\rlap{\lower4pt\hbox{$\sim$}}
    \raise1pt\hbox{$<$}}}                
\def\gsim{\mathrel{\rlap{\lower4pt\hbox{$\sim$}}
    \raise1pt\hbox{$>$}}}            
\def\thefootnote{\fnsymbol{footnote}}
\newcommand\beq{\begin{eqnarray}}
\newcommand\eeq{\end{eqnarray}}

\begin{document}

\renewcommand{\theequation}{\arabic{section}.\arabic{equation}}
\renewcommand{\thefigure}{\arabic{section}.\arabic{figure}}
\renewcommand{\thetable}{\arabic{section}.\arabic{table}}


\title{\Large \baselineskip=10pt  Probing the era of reheating for reconstructed inflationary potential in the \emph{RS $\rm II$} braneworld}

\vskip 2cm

\author{Sukannya Bhattacharya$^{1}$, Kumar Das$^{2}$, Mayukh~R.~Gangopadhyay$^{3}$}
\affiliation{
\it$^1$Theoretical Physics Division, Physical Research Laboratory, Navrangpura, Ahmedabad-380009, India.\\
\it$^2$S. N. Bose National Centre For Basic Sciences,  JD Block, Sector-III, Salt Lake City, Kolkata-700106, India.\\
\it$^3$Centre For Theoretical Physics, Jamia Millia Islamia, New Delhi-110025, India.}

\begin{abstract}\normalsize \baselineskip=10pt  
We analyse the epoch of reheating after an inflationary phase in the Randall Sundrum (RS) Type-$\rm II$ braneworld, where we did not consider any particular model of inflation, but rather reconstructed the inflationary potential solving the flow equations using Monte Carlo (MC) approach. It is shown numerically that a potential conceived  through the MC reconstruction technique can  be represented by an effective potential as a function of the number of e-foldings ($N$). Then, the epoch of reheating is studied for this reconstructed potential. The relation between the reheating temperature ($T_{\rm reh}$) and the 5-dimensional Planck mass ($M_5$) is established. Moreover, it is argued that there is a stringent bound on the critical reheating temperature that also translates to a tight bound on $M_5$ . 
\end{abstract}

\maketitle

\vspace{0.001in}


\baselineskip=15.4pt

\vspace{1cm}

\maketitle



\section{Introduction}\label{sec:intro}
\setcounter{equation}{0}
\setcounter{figure}{0}
\setcounter{table}{0}
\setcounter{footnote}{1}
Standard hot big bang cosmology is a very successful theory when combined with the inflationary paradigm. Inflationary epoch is required to solve some of the initial condition problems of this cosmological model, dubbed as the `Big bang puzzles'~\cite{Liddle:2000cg,Kolb:1990vq} . It was later realised that a phase of accelerated expansion (inflation) of the nascent universe not only solves the puzzles but also provides the seeds of primordial density fluctuations which plays the critical role in the formation of the large scale structures in the Universe. Several inflationary models have been proposed since the first model of inflation was initiated in~\cite{Guth:1980zm}. In the recent past, observations by the Wilkinson Microwave Anisotropy Probe (WMAP)~\cite{Hinshaw:2012aka} and the Planck mission~\cite{Ade:2015zua}  have constrained the inflationary era severely. In particular, the Planck 2015 inflation analysis \cite{Ade:2015zua} has  ruled out many of the theoretically popular models. 
In 2018, final results by the Planck mission is reported in \cite{Aghanim:2018eyx,Akrami:2018odb}, which constrain the models of inflation even more.

To find a viable model of inflation, there are number of reasons and ways to go beyond the standard model of cosmology as well as the standard model of particle physics. One of the problems which makes theoretical physicists really uncomfortable is known as the `hierarchy' problem. One way to define this problem is the difference of scales related to the fundamental forces in nature. One of the proposed solution to this problem is to introduce compact extra dimensions. But unfortunately, this creates a new hierarchy between the weak force and the compact extra dimensions. A possible solution was proposed by Lisa Randall and Raman Sundrum in~\cite{Randall:1999ee, Randall:1999vf}. In this model, the observed universe is embedded in a five-dimensional anti-de Sitter space ($AdS_5$). An interesting implication of this model when it is projected in the $(3+1)$ spacetime, is the modification of the Friedmann equation~\cite{Langlois:2002bb,Binetruy00, shiromizu, ida1, Maartens:2010ar}.\\
\begin{equation}
H^2 = \left(\frac{\dot{a}}{a}\right)^2
=\frac{8 \pi G_{\rm N}}{3} \rho
-\frac{K}{a^2}+\frac{\Lambda_{4}}{3}
+\frac{\kappa_{5}^4}{36}\rho^2 + \frac{\mu}{a^4} ~.
\label{Friedmann}
\end{equation}
Here, $H$ is the Hubble parameter, $a(t)$ is the scale factor at time $t$. $\rho$ is the standard 3-space matter energy density. $G_N$ is the four dimensional Newton's gravitational constant. The five-dimensional gravitational constatnt $\kappa_5$ is defined in terms of $G_N$ as,
\begin{equation}
\kappa_{5}^4= \frac{48 \pi G_{\rm N}}{ \tau } ~,
\end{equation}
where $\tau$ is the intrinsic brane tension. $\kappa_5^{2}= M_5^{-3}$, where  $M_5$ is the five-dimensional Planck mass. The $\Lambda_4$ in the third term is the standard cosmological constant. The five dimensional cosmological constant is related to $\Lambda_4$ as, 
\begin{equation}
\Lambda_{4} = \kappa_{5}^4 \tau^2 /12 + 3 \Lambda_{5}/4 ~.
\end{equation}
The standard Friedmann equation does not contain the fourth and the fifth terms of Eq. (\ref{Friedmann}). The fifth term scales as $a^{-4}$ with $\mu$ as a constant, thus coined as the `dark radiation'. 
 The evolution of such dark radiation term and its effects on the evolution of the Universe have been  considered in~\cite{Ichiki:2002eh, Sasankan:2016ixg, Sasankan:2017eqr}. 
 
In this work, our chief aim is to study the effect of the term proportional to $\rho^2$ in Eq.~(\ref{Friedmann}). So, first we restrict ourselves to the spatially flat FRW metric as the $(3 + 1)$ spacetime and this amounts to setting $K = 0$. Next we consider that the constant term $\Lambda_4 = 0$ (which is none else than the cosmological constant problem in braneworld scenario and we do not address it here) and finally after neglecting the dark radiation term from the gravitational Einstein equations, the
modified Friedmann equation on the brane takes the following simple form:
\begin{align}
H^2 = \frac{\rho}{3 M_{\text{pl}}^2} \bigg(1+\frac{\rho}{2\tau}\bigg) .
\end{align}
The quadratic density term $\rho^2$ is derived from the junction condition for $a(t)$ at the brane surface. The modified Friedmann equation at high energy, where the $\rho^2$  term dominates, makes the dynamics of the inflation different from the standard scenario \cite{Brax:2004xh}. 
We note that in the limit $\tau\to\infty$, we recover the standard four-dimensional general relativistic results. However, in the high energy limit \emph{i.e.} $\rho>>2\tau$ which essentially means $M_5 \leq M_{\text{pl}}$, the $\rho^2$ term in the modified Friedmann equation can play an important role in the early Universe.
After end of reheating, in the radiation-dominated epoch, this term scales down as $a^{-8}$ thus becoming insignificant very quickly. The effect of the presence of this term in case of early inflationary observables were studied previously in~\cite{Gangopadhyay:2016qqa, Maartens:1999hf, Goheer:2002bq, Bennai:2006th, Calcagni:2004bh, Tsujikawa:2003zd, Calcagni:2013lya, Neupane:2014vwa, Okada:2014eva, Okada:2015bra}. 

In this work, rather than concentrating on any particular model of inflation, the flow equation technique is implemented for the braneworld (for references please go through \cite{Ramirez:2004fb}). This approach was first introduced by Kinney \cite{Kinney:2002qn} to generate a large number of inflationary models through a random process compatible with the observation. The rest of the paper is organized as follows, in section~ \ref{sec:potrec}, we re-investigate the flow equation technique in the braneworld and the Monte Carlo (MC) simulated Hubble parameter is developed. After that, a comparison between the MC reconstructed potential with the attractor reconstruction is established in section~\ref{sec:sect}. Thereafter, in section~\ref{sec:reh}, we have restricted to the high energy limit of the braneworld and subsequently the constraints on the reheating temperature as well as the braneworld model parameter are studied. Finally in section~\ref{sec:conl} we have drawn the conclusion of our findings.


\section{Hubble Reconstruction}
\setcounter{equation}{0}
\setcounter{figure}{0}
\setcounter{table}{0}
\setcounter{footnote}{1}
\label{sec:potrec}
\begin{figure}[!htb]
    \centering
    \includegraphics[width=12.0cm]{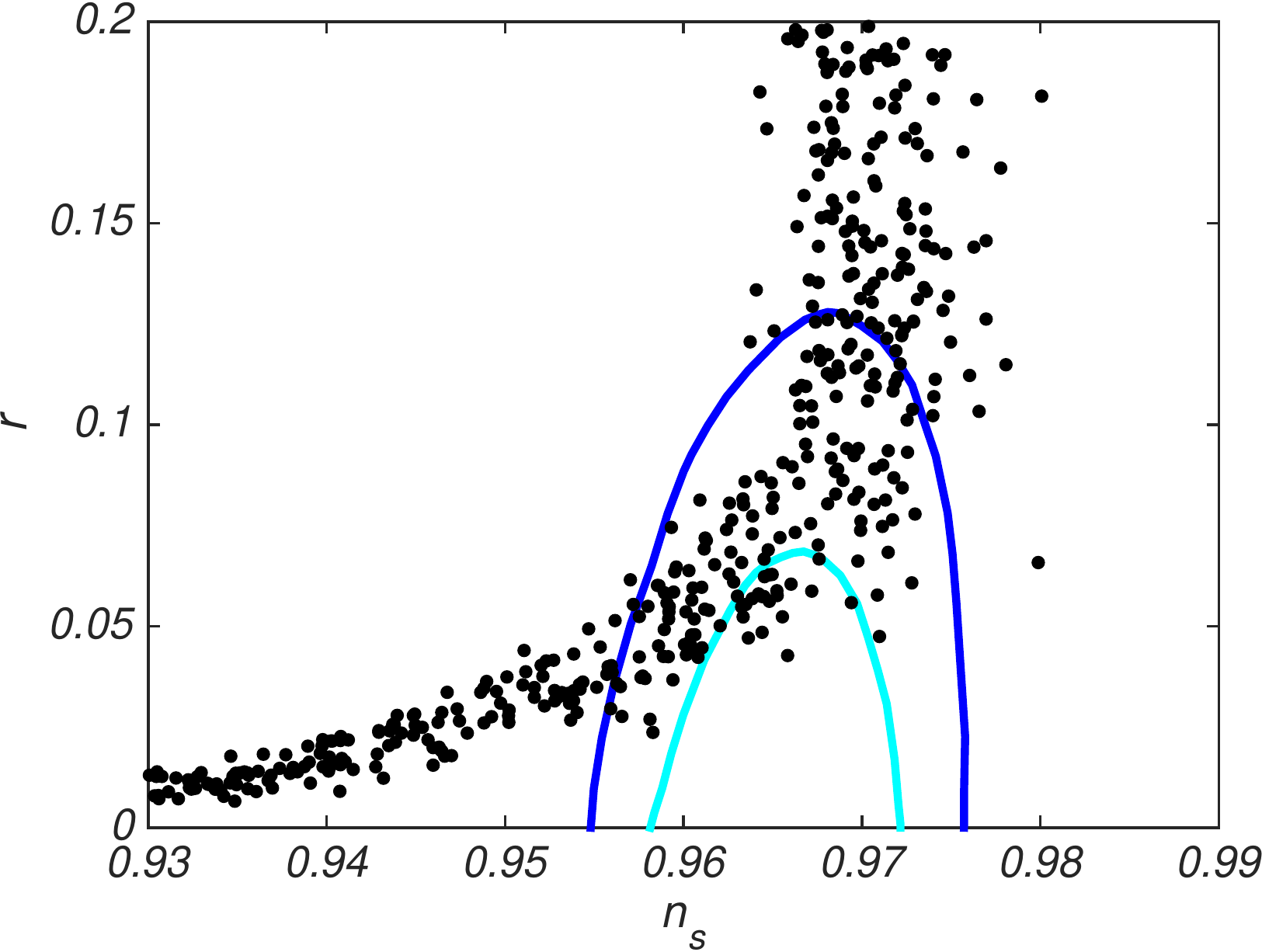} 
             \caption{\small The MC reconstructed values of spectral index ($n_s$) and tensor-to-scalar ratio ($r$)  are plotted (black dots) for different values of the number of e-folds in the range 50 to 74. The light blue and the dark blue contours represent the most stringent Planck 2018 1-$\sigma$ and 2-$\sigma$ confidence limits. ($\sigma$ here is the standard deviation of the posterior probability distribution of the cosmological parameters $n_s$ and $r$).} 
\label{mcmc_nsr}
\end{figure} 

There exists different reconstruction schemes for obtaining the shape and form of the inflationary potential. Out of them, Monte Carlo (MC) reconstruction is a technique which is based on a stochastic method to invert the observational constraints and thereby yielding the inflationary potential compatible with observation~\cite{Kinney:2002qn,Easther:2002rw}. This reconstruction method generates the potential numerically instead of giving an exact closed form expression. In this section, we will discuss in brief about the MC reconstruction process. Later in section~\ref{sec:sect}, we will show that for the braneworld at high energy limit, the reconstructed potential via MC method is a very good approximation to the inflaton potential obtained through attractor reconstruction~\cite{Chiba:2015zpa}.
\begin{figure}[!htb]
    \centering
    \includegraphics[width=11.0cm]{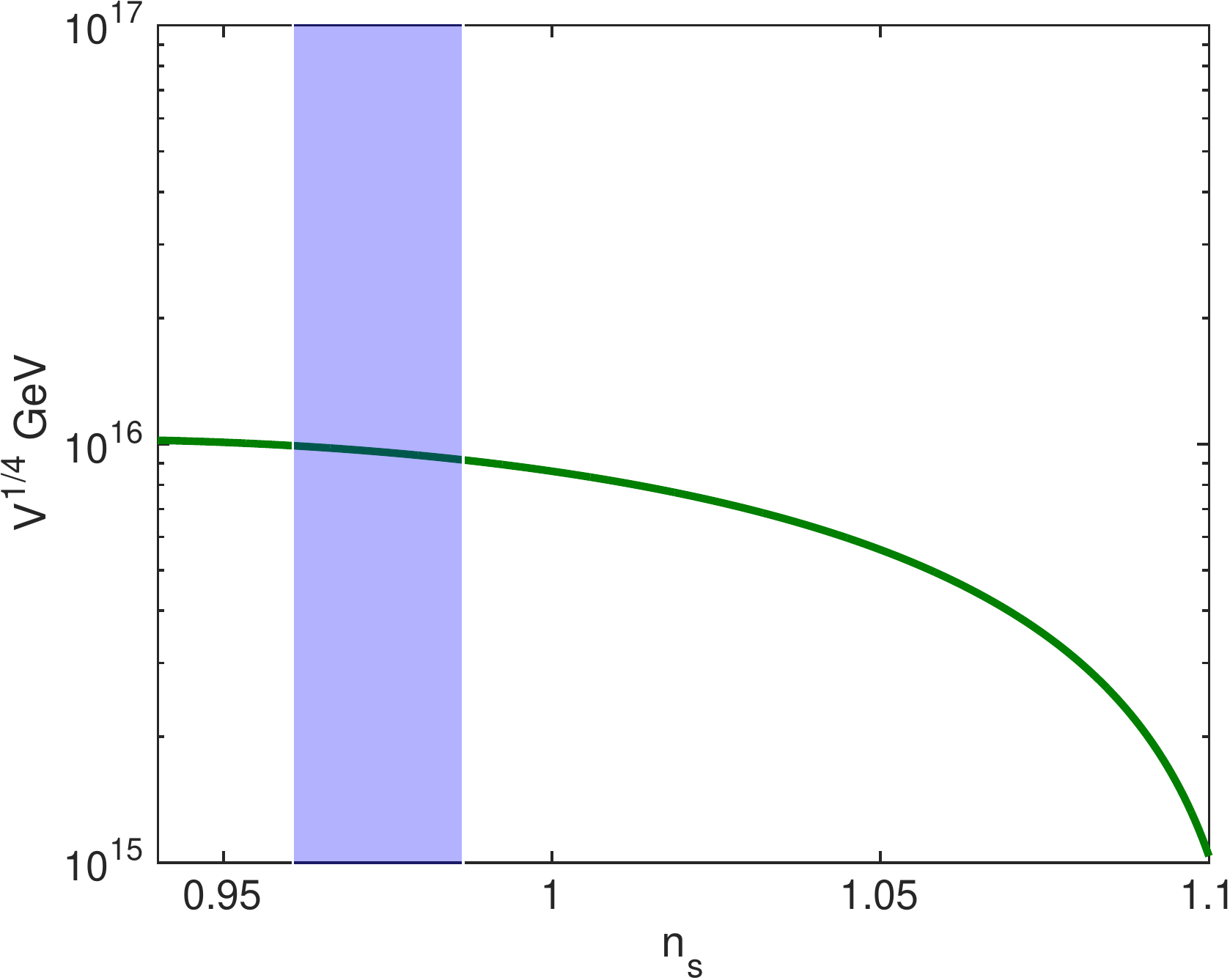} 
             \caption{\small The green line represent the best-fit to the MC generated values of Hubble during inflation with the scalar spectral index ($n_s$).The two vertical lines represent the Planck 2018 1-$\sigma$ bound on $n_s$.} 
\label{nsh}
\end{figure} 
The MC reconstruction starts with a set of flow equations, defined in general in terms of an infinite hierarchy of Hubble slow roll parameters. 
These flow equations, governing the MC reconstruction, are identically a set of coupled differential equations which completely specify the background cosmological evolution. In the standard cosmology, following Kinney~\cite{Kinney:2002qn}, these parameters involving in the flow equations are defined as --- 
\begin{equation}
\epsilon _H(\phi)=\frac{M_{\text{pl}}^{2}}{4\pi}\bigg(\frac{H'(\phi)}{H(\phi)}\bigg)^2~,
\end{equation}
\begin{equation}
^l\lambda_H=~\bigg(\frac{M_{\text{pl}}^{2}}{4\pi}\bigg)^l\frac{H'(\phi)^{l-1}}{H(\phi)^{l}}\bigg(\frac{d^{(l+1)}H(\phi)}{d\phi^{(l+1)}}\bigg);~l\geq 1,
\end{equation}
where prime represents the derivatives with respect to the inflaton field $\phi$. The index $l$ represents the order of the slow roll parameters and can go up to infinity.
From these definitions, one gets the flow equations for the standard inflation as --- 
\begin{equation}
\frac{d\epsilon _H}{dN}=~ \epsilon _H[\sigma _H+ 2\epsilon _H] ~,
\end{equation}
\begin{equation}
\frac{d\sigma _H}{dN}=~2^{2}\lambda_H- 5\sigma _H \epsilon _H- 12\epsilon _H^2 ~,
\end{equation}
\begin{equation}
\frac{d(^l\lambda_H)}{dN}=~\bigg[\bigg(\frac{l-1}{2}\bigg)\sigma _H+ (l-2)\epsilon _H\bigg](^l\lambda_H)+~^{l+1}\lambda_H ~ .
\end{equation}
To review the flow equation reconstruction process in details, the reader is suggested to refer to~\cite{Kinney:2002qn}. Using this method, flow equations are derived in~\cite{Ramirez:2004fb} for the RS $\rm II$ braneworld. For this case, a new set of slow-roll parameters can be defined for the high-energy  regime --- 
\begin{eqnarray}
\epsilon _H &=& \frac{M_5^3}{4\pi}\frac{H'^2}{H^3}, \\
\eta _H &=& \frac{M_5^3}{4\pi}\frac{H''}{H^2}-\epsilon _H {~ ~ \rm and }\\
^l\lambda_H &=& \bigg(\frac{M_5^3}{4\pi}\bigg)^l\frac{H'(\phi)^{l-1}}{H(\phi)^{2l}}\bigg(\frac{d^{(l+1)}H(\phi)}{d\phi^{(l+1)}}\bigg);~l\geq 1.
\end{eqnarray}
Therefore, the set of flow equations is now modified to --- 
\begin{eqnarray}
\frac{d\epsilon _H}{dN} &=& \epsilon _H (\sigma _H+3\epsilon _H),\label{RSfloweq1}\\
\frac{d\sigma _H}{dN} &=& -8\epsilon _H \sigma _H -30\epsilon ^2_H+2(^{2}\lambda_H) {~ ~ \rm and }\label{RSfloweq2}\\
\frac{d(^l\lambda_H)}{dN} &=& \bigg[\bigg(\frac{l-1}{2}\bigg)\sigma _H+ (l-3)\epsilon _H\bigg](^l\lambda_H)+~^{l+1}\lambda_H,
\label{RSfloweq3}
\end{eqnarray}
where $\sigma _H \equiv 2\eta _H - 4\epsilon _H$.

Using Eq.s~\eqref{RSfloweq1} - \eqref{RSfloweq3}, we can now reconstruct the inflationary observables-- the scalar spectral index ($n_s$) and the tensor-to-scalar ratio ($r$) from flow equations of slow roll parameters upto $l= 6$. The outcome is shown in the Fig.~\ref{mcmc_nsr}.

Fig.~\ref{nsh} represents the change of Hubble parameter, $H$, reconstructed through the flow equations corresponding to the modified Friedmann equation in the braneworld. The change of $H$ with respect to $n_s$ shows that a slow roll inflation is compatible with observational data only for a red tilted spectrum. The polynomial fitting carried out here clearly indicates the $n_s$ value is less than 1 for the relevant scales in CMB.

\section{Comparison between MC and attractor reconstruction}
\setcounter{equation}{0}
\setcounter{figure}{0}
\setcounter{table}{0}
\setcounter{footnote}{1}
\label{sec:sect}
The previous section dealt with the MC reconstruction of the Hubble parameter which is completely a numerical technique. However, there exists a direct reconstruction procedure, where the reconstruction of the potential is done from a parametrization of either the spectral index $(n_s)$ or the tensor-to-scalar ratio $(r)$ as a function of the e-folds $N$. In this reconstruction scheme, the form of the inflaton potential is obtained via two steps as highlighted in \cite{Chiba:2015zpa}. First, one obtains the scalar potential as a function of e-folds, \emph{i.e.} $V(N)$ and subsequently, using the relation between $N$ and the inflaton field $\phi$, one gets $V(\phi)$. The corresponding scheme for the braneworld inflation scenario is thoroughly described in \cite{Herrera:2019xhs, myrzakul}. For braneworld inflation, the reconstruction can be done analytically if one considers the high energy limit \emph{i.e.} $V>>\tau$ .

In this section our aim is to show that the MC reconstructed potential  is a very good approximation to the potential obtained from a  parameterization of $n_s$ or $r$ in terms of $N$. 
To begin with, we consider the usual potential slow roll parameters for the case of braneworld inflation 
~\cite{Herrera:2019xhs}: 
\begin{eqnarray}
\epsilon_V  =  \frac{1}{2\kappa}\bigg(\frac{V'}{V}\bigg)^2 \frac{(1+\frac{V}{\tau})}{(1+\frac{V}{2\tau})^2} \label{epsv_brane} 
 ~~ {\rm and} ~~
\eta_V  =  \frac{1}{\kappa}\frac{V''}{V(1+\frac{V}{2\tau})}, \label{etav_brane}
\end{eqnarray}
where again prime denotes derivative with respect to the inflaton field $\phi$ and $\kappa=8\pi G_{\rm N}= 1/M_{\text{pl}}^2$. Under the slow roll approximation, the power spectrum for curvature perturbations is
\begin{equation}
P_{\mathcal{R}} \equiv A_s \bigg(\frac{k}{k_0}\bigg)^{n_s-1}=\bigg(\frac{H^2}{\dot{\phi}^2}\bigg)\bigg(\frac{H^2}{2\pi}\bigg)\simeq
 \frac{\kappa^3}{12\pi^2}\frac{V^3}{V'^2}\bigg(1+\frac{V}{2\tau}\bigg)^3~,
 \label{power_brane} 
\end{equation}
where, $A_s$ is the amplitude of the power spectrum at the pivot scale $k_0$. The scalar spectral index $n_s-1=\frac{d\ln P_{\mathcal{R}}}{d\ln k}$ can then be written in terms of $\epsilon_V$ and $\eta_V$ as
\begin{equation}
n_s-1=-6\epsilon_V+2\eta_V. \label{ns_brane}
\end{equation}

The number of e-folds passed from a particular time $t$ until the end of inflation $t_{\text{end}}$ is defined with the slow roll assumption as well:
\begin{equation}
N=\int_t^{t_{\text{end}}} H dt = \kappa \int ^{\phi}_{\phi_{\text{end}}} \frac{V}{V'}\bigg(1+\frac{V}{2\tau}\bigg)d\phi , \label{efold_brane}
\end{equation}
where $\phi_{\text{end}}$ corresponds to the end of inflation and $N=N_k$ at the pivot scale $k=0.002~{\rm Mpc^{-1}}$. Now, it is convenient to use $N$ instead of $\phi$ as the dynamical variable such that, 
\begin{equation}
\frac{dN}{d\phi}=\sqrt{\frac{\kappa V}{V_{,N}}}\sqrt{\bigg(1+\frac{V}{2\tau}\bigg)},\label{dNdphi_brane}
\end{equation}
where $V_{, N}=\frac{dV}{dN}$. Then $V'^2=\kappa V\bigg(1+\frac{V}{2\tau}\bigg)V_{,N}$ and the scalar spectral index from Eq.~\eqref{ns_brane} is:
\begin{equation}
n_s-1=\frac{V_{,NN}}{V_{,N}}-2\frac{(1+\frac{V}{\tau})}{V(1+\frac{V}{2\tau})}V_{,N}=\bigg[\ln \bigg[ \frac{V_{,N}}{V^2(1+\frac{V}{2\tau})^2}\bigg]\bigg]_{,N}.\label{nsN_brane}
\end{equation}
Now, in the high energy limit, $V\gg \tau$ and therefore from~\cite{Herrera:2019xhs}, 
\begin{equation}
n_s-1=\bigg[\ln \bigg[ \frac{V_{,N}}{V^4}\bigg]\bigg]_{,N}.\label{nsN_brane_HE}
\end{equation}
Since we have obtained the exact values of $n_s$ for a few values of $N$ from the MC  
reconstruction in the last section, therefore the above equation can be integrated keeping $n_s$ at its numerical value. The resulting potential is:
\begin{equation}
V=\bigg(3c_1\frac{\exp [(n_s-1)N]}{1-n_s}+c_2\bigg)^{-1/3},\label{recpot1HE}
\end{equation}
where $c_1$ and $c_2$ are the constants of integration. Again, from Eq.~\eqref{power_brane}, 
\begin{equation}
P_{\mathcal{R}}=\frac{\kappa^2}{48\pi^2\tau^2}\bigg(\frac{V^4}{V_{,N}}\bigg)=\frac{\kappa^2}{48\pi^2\tau^2c_1}\exp[-(n_s-1)N].\label{c1_brane}
\end{equation}
Therefore, writing $c_1$ in terms of $N_k$ allows us to write the reconstructed potential at pivot completely in terms of $n_s$, considering the value of the power spectrum to remain the same for few e-folds around pivot $N_k$.:
\begin{equation}
V=\bigg[\frac{\kappa^2}{16\pi^2\tau^2P_{\mathcal{R}}}\frac{1}{(1-n_s)}+c_2\bigg]^{-1/3}.\label{recpot2_HE}
\end{equation}
We note that this potential is very similar in form to the reconstructed potential obtained in~\cite{Herrera:2019xhs},
\begin{align}
V(N) = 3^{-1/3}\bigg(\frac{\alpha}{N} + \beta \bigg)^{-1/3} , \label{vnpot}
\end{align}

where $n_s=1-2/{N}$ is used to integrate Eq.~\eqref{nsN_brane_HE}. Additionally, we know that this potential, for some regime in the parameter space of $\alpha$ and $\beta$, has an attractor like feature because it is consistent with the consistency relation $r\propto(n_s -1)^2$, which is a typical signature of attractor type models of inflation (see also Sec.~\ref{sec:reh}). The range of parameters for which this potential is an attractor, subject to its derivation from a parameterization of $n_s(N)$, is same even if we start from a parameterization $r\propto1/N^2$ instead of parameterizing $n_s$.

Of the two integration constants, $c_1$ is determined here in Eq.~\eqref{c1_brane} in terms of the observable value of the scalar power spectrum from Planck in a similar way as the integration constant $\alpha=\frac{\kappa ^2}{48\pi^2\tau^2}\frac{N^2}{P_{\mathcal{R}}}$ is determined in~\cite{Herrera:2019xhs}. On the other hand, $c_2$ and $\beta$ are arbitrary constants in the respective cases and therefore one of them can be expressed in terms of the other so that the reconstructed potential simulates the exact attractor behavior in Eq.~\eqref{vnpot}. For the general reconstructed potential~\ref{recpot2_HE} as an attractor,
\begin{equation}
c_2=\frac{\kappa^2}{16\pi^2\tau^2P_{\mathcal{R}}(1-n_s)}\bigg(1+\frac{4\beta}{\alpha (1-n_s)}\bigg). \label{c2betaalpha}
\end{equation}
It is interesting to note here that for the value $\beta=0$, $c_2\neq 0$ becomes an independent constant of integration, to be determined solely from the brane tension $\tau$. Moreover, the second term in the parenthesis starts contributing to the value of $c_2$ only when $4\beta/\alpha \sim (1-n_s) \sim 0.0351$ (Planck 2018).\\
In the high energy limit of braneworld, the tensor-to-scalar ratio is given as 
\begin{figure}[t]
    \centering
    \includegraphics[width=11.0cm]{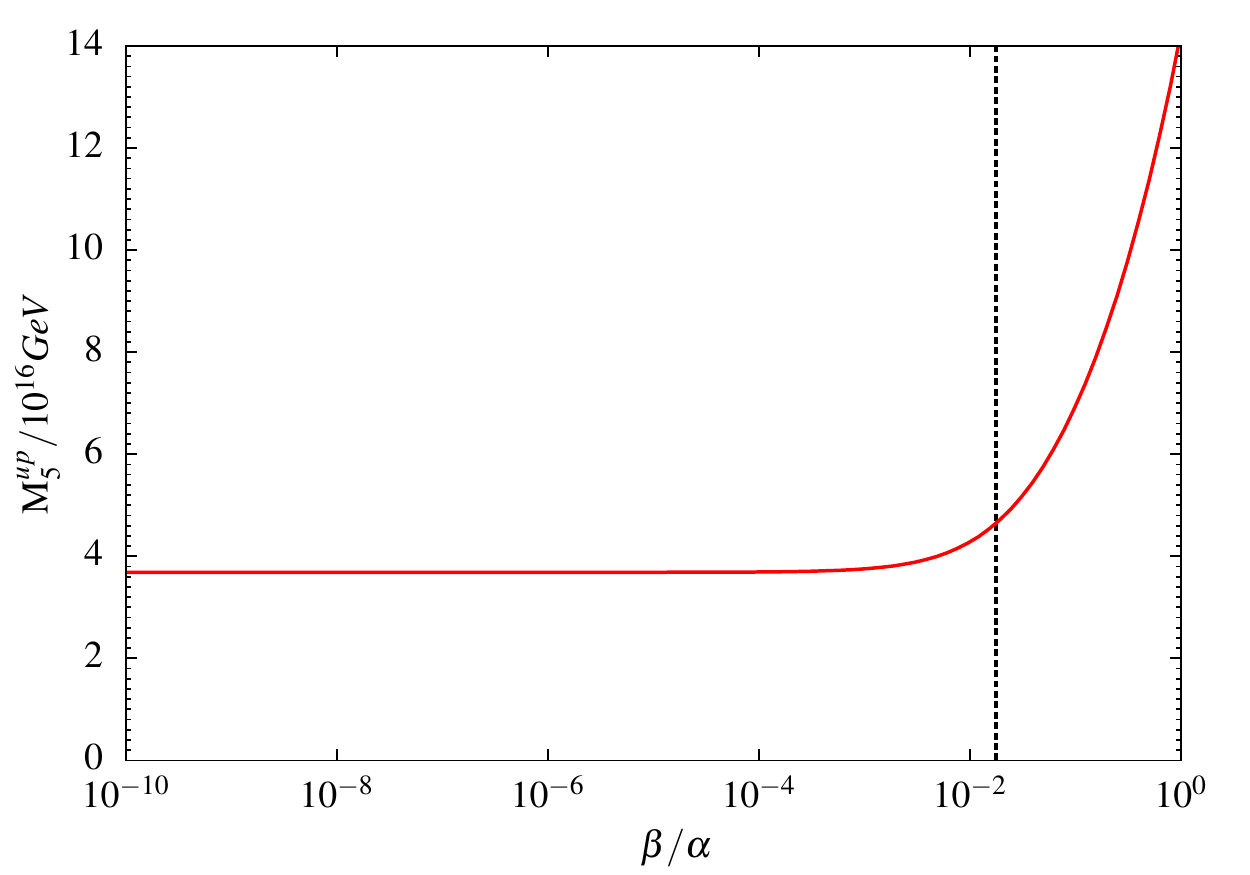} 
             \caption{\small The red curve shows the variation of the upper limit on $M_5$ with different values of the constants of integration $\beta$ and $\alpha$. This limit comes from the constraint $r \leq 0.064$, Planck 2018 in Eq.~\eqref{mfr}. The black dashed line marks the value $2\beta/\alpha\sim (1-n_s)$ (with $n_s=0.9649$, Planck 2018), above which the contribution from $\beta/\alpha$ becomes relevant in $M_5$. } 
\label{mrg}
\end{figure} 
\begin{equation}
r=48\tau\bigg(\frac{V_{,N}}{V^2}\bigg).\label{rHE_brane}
\end{equation}
So, for the potential in Eq.~\eqref{recpot1HE},
\begin{eqnarray}
r&=&48\tau c_1 V^2 \exp[(n_s-1)N]\nonumber \\
&=&\frac{\kappa^2}{\pi^2P_{\mathcal{R}}\tau}\bigg[\frac{\kappa^2}{16\pi^2\tau^2P_{\mathcal{R}}}\bigg(\frac{2}{1-n_s}+\frac{4\beta/\alpha}{(1-n_s)^2}\bigg)\bigg]^{-2/3},\label{rMCHE_brane}
\end{eqnarray}
where, in the second expression, $c_1$ is written from Eq.~\eqref{c1_brane} and $c_2$ from Eq.~\eqref{c2betaalpha}. Now, considering the Planck 2018 best-fit values $P_{\mathcal{R}}=2.2\times10^{-9}$ and $n_s=0.9649$ in Eq.~\eqref{rMCHE_brane}, the upper bound on $r\leq0.064$ gives an upper bound on the brane tension $\tau$ for each value of $\beta/\alpha$. For $\beta=0$, this gives: $\tau \leq 7 \times 10^{-11} M_{\rm Pl}^4$. The contribution from the second term in Eq.~\eqref{rMCHE_brane} is relevant only above $2\beta/\alpha\sim (1-n_s)$, and below that, the upper limit on $\tau$ remains the same as $\beta=0$ case. Therefore, the upper limit on $r$ from Planck 2018 constrains $M_5=\bigg(\frac{\tau}{6\kappa}\bigg)^{1/6}$ from above as $M_5 \lesssim 2\times 10^{-2}M_{\rm Pl} $ for $\beta=0$, whereas for nonzero values of $\beta$, the limit is:
\begin{equation}
M_5 \leq \frac{r_{\rm max}^{1/2}P_{\mathcal{R}}^{1/6}\pi ^{1/3}}{16^{1/3}6^{1/6}\kappa ^{1/2}}\bigg[\frac{2(1-n_s)+4\beta/\alpha}{(1-n_s)^2}\bigg]^{1/3}.
\label{mfr}
\end{equation}
Fig.~\ref{mrg} shows the variation on the upper limit on $M_5$ with $\beta/\alpha$.

An interesting point to note here is that, by construction, this limit on $M_5$ is independent of the number of e-folds $N$ when the integration constant $c_1$ is written in terms of the observables in CMB and $c_2$ depends on the integration constants $\alpha$ and $\beta$ (the integration constants related to the attractor reconstruction).
 
\begin{figure}[!htb]
    \centering
    \includegraphics[width=11.0cm]{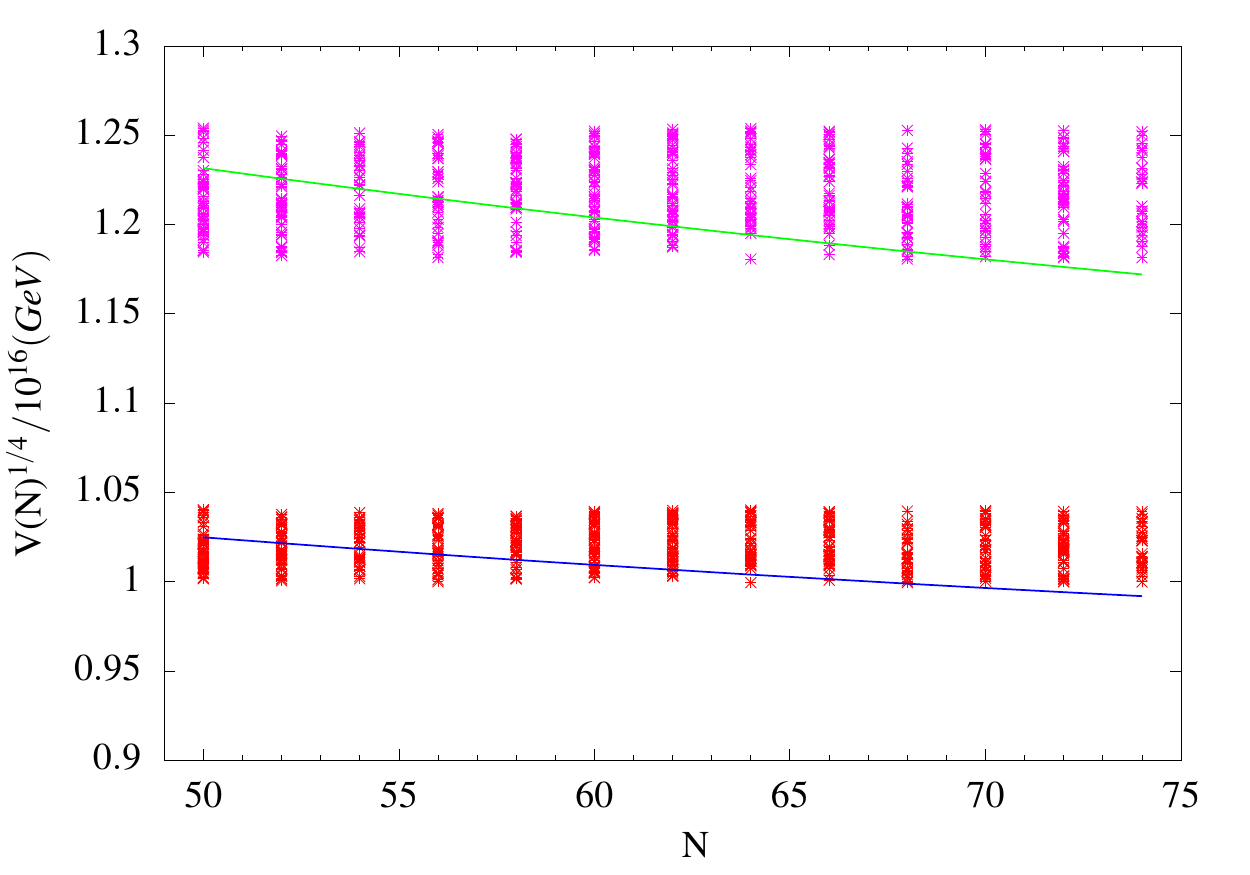} 
             \caption{\small This plot shows the MC reconstructed potential (as points) from Eq.~\eqref{recpot2_HE} and the attractor potential (as solid lines) from Eq.~\eqref{vnpot} as functions of the number of e-folds $N$. All the $n_s$ values that are considered to find the reconstructed potential (red and magenta points) comply with the $2\sigma$ bound from Planck 2018. The red points and blue line represent $\beta/\alpha = 10^{-10}$ case, for which $\tau$ is kept at its maximum allowed value $\tau=7 \times 10^{-11}M_{\rm pl}^{4}$. The magenta points and green line represent $2\beta/\alpha = (1-n_s\vert_{\rm best fit,Planck 2018})$ case, for which $\tau$ is kept at its maximum allowed value $\tau=2.9 \times 10^{-10}M_{\rm pl}^{4}$. }
\label{fig_c}
\end{figure} 
The comparison between the forms of the reconstructed potentials through the MC reconstruction technique in Eq.~\eqref{recpot2_HE} and of an attractor in Eq.~\eqref{vnpot} is particularly necessary for understanding the reheating dynamics. The reason is that the MC technique reconstructs the potential for only very few e-folds near the pivot scale and therefore is not reliable to provide precise values for the quantities at the end of inflation. Fig.~\ref{fig_c} shows how close the reconstructed potential in Eq.~\eqref{recpot2_HE} is to an attractor for a range of e-folds near the pivot for two different $\beta/\alpha$ and their corresponding maximum values of $\tau$ from Eq.~\eqref{rMCHE_brane}. Fig.~\ref{fig_c} also shows that the spread of the MC reconstructed points, and therefore the uncertainty in the reconstructed potential is more for $2\beta/\alpha=(1-n_s\vert_{\rm best fit,Planck 2018})$ than that for $\beta/\alpha=0$ case. This uncertainty may lead to the deviation of the reconstructed potential from a plateau behaviour. 
Similarly, The attractor potential (solid lines) also show more slope for the case when the contribution of the second term in Eq.~\eqref{vnpot} becomes relevant. Throughout the analysis of this section, the inequality $M_{\rm pl}>M_5>V_{\rm inf}^{1/4}$ is satisfied, even for large values of $\beta/\alpha$, as is evident from the $M_5$ axis in Fig.~\ref{mrg} and the $V_{\rm inf}^{1/4}$ axis in Fig.~\ref{fig_c}.



\section{Reheating Analysis}
\setcounter{equation}{0}
\setcounter{figure}{0}
\setcounter{table}{0}
\setcounter{footnote}{1}
\label{sec:reh}
In the previous section, the MC reconstructed potential is established to be very similar to the effective inflationary potential for attractor as a function of e-folds~\cite{Herrera:2019xhs}. Therefore, for the rest of the paper, we will use the attractor form of the potential given below: 
\begin{align}
V(N_k) = 3^{-1/3}\bigg(\frac{\alpha}{N_k} + \beta \bigg)^{-1/3} , \label{vnpot2}
\end{align}
This potential has two interesting regimes for the parameter $\beta$. For $\beta=0$, the reconstructed effective potential becomes a monomial potential. On the other hand, for $\beta\ne 0$, we get either an exponentially flat potential or a monomial potential depending on whether the quantity $\alpha/\beta< N_k$ or $>N_k$~\cite{Herrera:2019xhs}.  In this section, we consider the regime $\beta\ne 0$ and $\alpha/\beta < N_k$, when the reconstructed potential is a plateau or an attractor type model.

Now, after the end of inflation, there will be a period of reheating when inflaton energy density is dumped into a thermal bath of  relativistic particles, characterised by the reheating temperature ($T_{\text{reh}}$). Once the form of $V(N_k)$ is known, we can relate this reheating temperature with $N_k$ for small value of e-folds around the pivot value~\cite{Dai:2014jja,Munoz:2014eqa}. For large values in the number of e-folds, $N_k$, the reheating temperature, $T_{\text{reh}}$, will have exponential sensitivity with $N_k$ as will be seen shortly. Hence $T_{\text{reh}}$ will be pushed to higher values close to the inflationary energy density (see also \cite{Kuroyanagi:2014qaa} for a details on the reheating with non-minimal coupling). 

To begin our analysis, we consider a particular mode with wave number $k$. When this mode exited the horizon, the co-moving Hubble scale, $a_k H_k$, pertaining to that mode is related to the same of present time \emph{i.e.}, $a_0 H_0$, as, 
\begin{align}
\frac{k}{a_0 H_0} = \frac{a_k}{a_{\text{end}}}\frac{a_{\text{end}}}{a_{\text{reh}}}\frac{a_{\text{reh}}}{a_0}\frac{H_k}{H_0} , \label{remode}
\end{align} \\
where $a_{\text{end}}$ and $a_{\text{reh}}$ are the scale factor at end of inflation and during reheating respectively. Also, $a_0$ and $H_0$ are the present value of the scale factor and the Hubble parameter respectively and $H_0 = 2.133 h \times 10^{-42}$ GeV, where $h=0.67$ is the dimensionless Hubble parameter. Now using the definition of inflationary e-folds,
$e^{-N_k}=\frac{a_k}{a_{\text{end}}}$ in eqn.~\eqref{remode} we obtain \\ 
\begin{figure}[t]
    \centering
    \includegraphics[width=12.0cm]{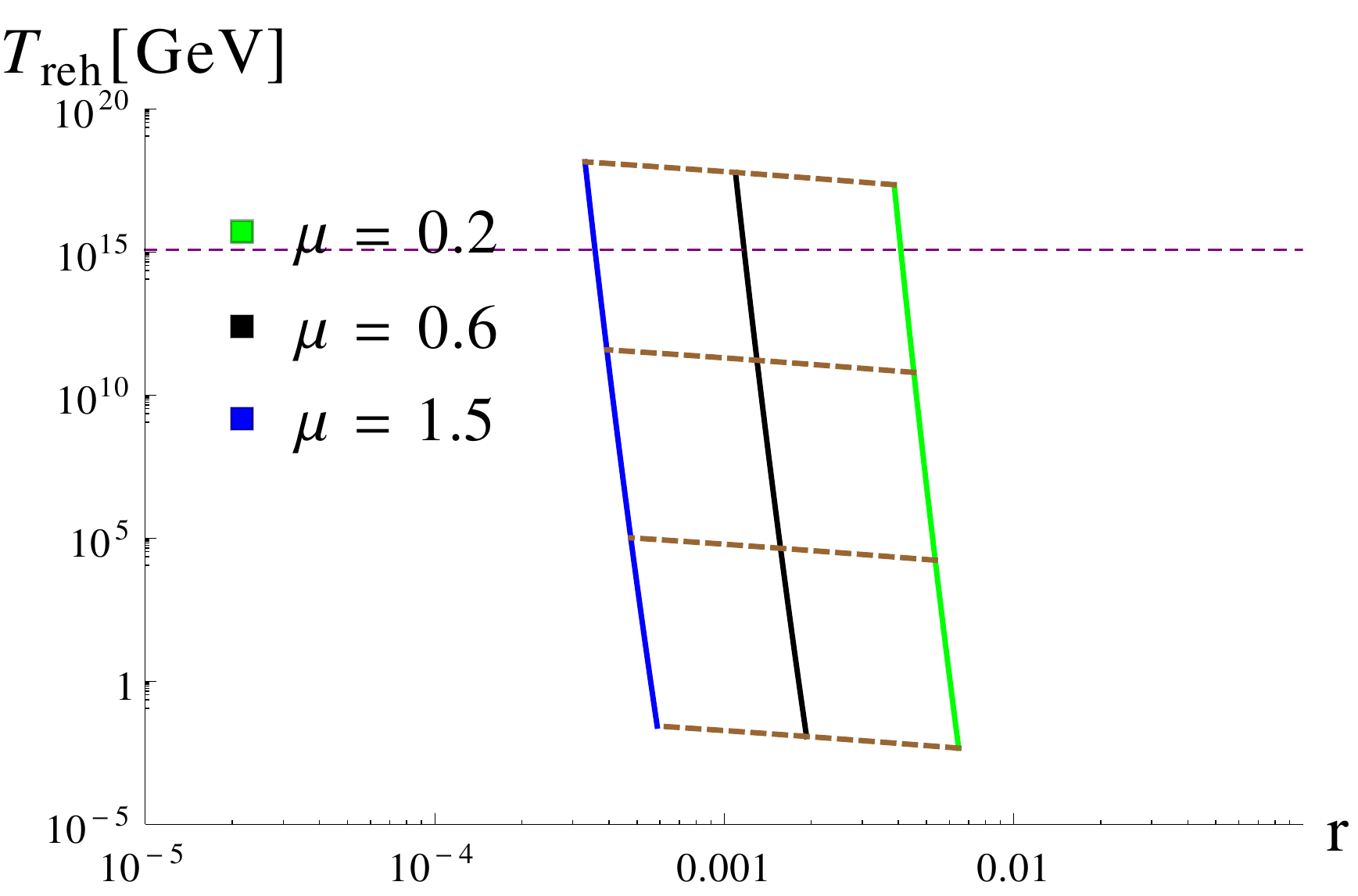} 
             \caption{\small Plot of $T_{\text{reh}}$ (at $k=0.002$ Mpc$^{-1}$) as a function of tensor-to-scalar ratio ($r$) for $\mu=0.2$ (green), $\mu=0.6$ (black) and $\mu=1.5$ (blue) respectively from left to right. The dashed brown lines correspond to the variation of $T_{\text{reh}}$ for $N_k =45,50,55$ and $60$ from bottom to top. The critical reheat temperature ($T_{\text{reh}}^{\text{cr}}$) is shown by the dashed purple line.} 
\label{reha_swamp}
\end{figure}
\begin{align}
\text{ln} ~ \bigg(\frac{k}{a_0 H_0}\bigg) = - N_k + \text{ln} ~ \frac{a_{\text{end}}}{a_{\text{reh}}} + \text{ln} ~ \frac{a_{\text{reh}}}{a_0} + \text{ln} ~ \frac{H_k}{H_0} 
\end{align}
Assuming entropy conservation, $T_{\text{reh}}$ can be related to the CMB temperature $T_0$ as \cite{Dai:2014jja,Munoz:2014eqa}
\begin{align}
g_{s,reh} a_{\text{reh}}^3 T_{\text{reh}}^3 = a_0^{\,3} \bigg( 2T_{0}^{\,3} + \frac{21}{4} T_{\nu 0}^{\,3} \bigg) 
\end{align}
where $g_{\text{s,reh}}$ is the effective number of relativistic degrees of freedom for entropy at reheating and $T_{\nu\,0}$ is the current neutrino temperature.
 But we know that $T_{\nu\,0} = \big(\frac{4}{11} \big)^{1/3} T_0$. Therefore,
\begin{align}
\frac{a_{\text{reh}}}{a_0} = \bigg( \frac{43}{11 g_{\text{s,reh}}} \bigg)^{1/3} \times \frac{T_0}{T_{\text{reh}}}
\end{align}
However, the energy density at the end of reheating is $\rho_{\text{reh}} = \frac{\pi^2}{30} g_{\star,\text{reh}} T_{\text{reh}}^4$, where $g_{\star,\text{reh}}$ is the effective number of relativistic degrees of freedom at the end of reheating. Also, for simplicity, if we assume the standard canonical reheating scenario when the effective equation of state parameter ($w_{\rm re}$) is matter like, then 
\begin{align}
N_k = 67.35 - \text{ln}~\bigg(\frac{k}{a_0 H_0}\bigg) - \frac 1 3 ~ \text{ln}~{\frac{\rho_{\text{end}}}{\rho_{\text{inf}}}} + \frac 1 3 ~\text{ln}\,\bigg(\frac{g_{\star,\text{reh}}}{g_{\text{s,reh}}}\bigg) + \frac 1 3 ~ \text{ln} ~ \frac{T_{\text{reh}}}{H_k}  -  
 \frac 4 3 ~ \text{ln}\,\bigg(\frac{V_{\text{inf}}^{1/4}}{H_k}\bigg) .
\label{reh_eq}
\end{align} \\
In what follows, we shall use the above formula to calculate the reheating temperature. But before that, let us express $H_k$ in terms of the normalized tensor amplitude $r$. 
For braneworld in high energy regime $V\gg\tau$, the tensor-to-scalar ratio is $r ={A_t}/{A_s} =\frac{8}{A_s M_{\text{pl}}^2}\big(\frac{H_k}{2\pi}\big)^2 F^2(y) $, where $y=H_k M_{\text{pl}}\sqrt{\frac{6}{\tau}}$ and the function $F^2(y)\equiv \frac{3}{2} y $ \cite{copeland}. From this relation we can express $H_k$  as  
\begin{align}
H_k = \bigg(\frac{\pi^2}{6} r A_s M_{\text{pl}}\sqrt{\frac{2\tau}{3}}\bigg)^{1/3}, \label{hubb_brane}
\end{align}
where the amplitude of scalar perturbation is $\text{ln}(10^{10} A_s) = 3.040\pm{0.016}$ (taken from the current Planck 2018 TT + low E data \cite{Aghanim:2018eyx,Akrami:2018odb}). 

Now let us evaluate the ratio between the end energy density ($\rho_{\rm end}$) to the inflationary energy density ($\rho_{\rm inf}$). This can be found in the following way as
\begin{align}
\frac{\rho_{\text{end}}}{\rho_{\text{inf}}} \simeq \frac{V_{\text{end}}}{V_{\text{inf}}}\bigg( 1 + \frac{\dot{\phi}^2}{2V_{\text{end}}} \bigg) .
\end{align}
Exploiting the slow roll equation $3H\dot{\phi} + V_{\phi} \simeq 0 $, we found 
\begin{align}
\frac{\dot{\phi}^2}{2V} \simeq \frac 1 3 \frac{1}{2\kappa}\bigg(\frac{V_{\phi}}{V}\bigg)^2\frac{1}{1+V/{2\tau}} \label{keperatio}
\end{align} \\
\begin{figure}[t]
    \centering
    \includegraphics[width=13.1cm]{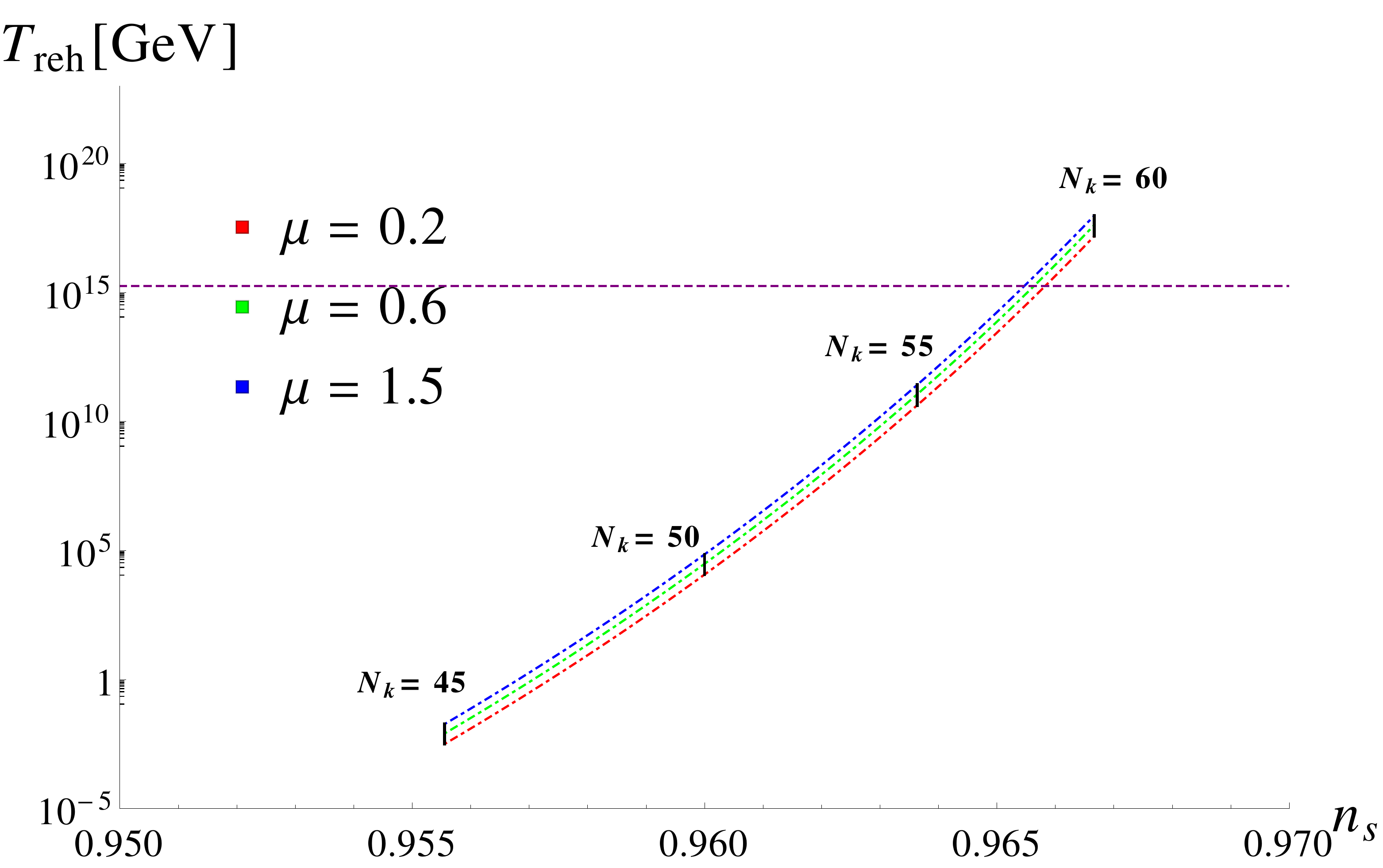} 
             \caption{\small Plot of $T_{\text{reh}}$ (at $k=0.002$ Mpc$^{-1}$) as a function of the scalar spectral index ($n_s$) for $\mu=0.2$ (red), $\mu=0.6$ (green) and $\mu=1.5$ (blue) respectively from left to right. The dashed brown lines show the variation of $T_{\text{reh}}$ for $N_k =45,50,55$ and $60$ from bottom to top. The critical reheat temperature ($T_{\text{reh}}^{\text{cr}}$) is shown by the dashed purple line.} 
\label{ns_variation}
\end{figure} 
At the high energy limit of the braneworld, when $V \gg \tau$, eqn.~\eqref{epsv_brane} and eqn.~\eqref{keperatio} yields
\begin{align}
\epsilon & \simeq \frac{1}{2\kappa} \bigg(\frac{V_{\phi}}{V}\bigg)^2 ~ \frac{4\tau}{V}~, \\
1+\frac{\dot{\phi}^2}{2V} & \simeq  1 + \frac 1 6 \epsilon_{0}~,
\end{align}
where at the end of inflation, $\epsilon \equiv \epsilon_{0}= 1$. Therefore,
\begin{align}
\frac{\rho_{\text{end}}}{\rho_{\text{inf}}(N_k)} \simeq \frac 7 6 \frac{V_{\text{end}}}{V_{\text{inf}}(N_k)} \label{rho_ra}
\end{align}
The next task is to get $V_{\rm inf}$ as a function of $N_k$ from eqn.~\eqref{vnpot2} and $V_{\rm end}$ from the condition $\epsilon_{0} =1$ and plug them in the r.h.s. of eqn.~\eqref{rho_ra}. After few steps of algebraic calculations, we found
\begin{align}
\frac{V_{\text{end}}}{V_{\text{inf}}(N_k)} =f(x) ~ \bigg(\frac{1+\mu^2 N_k}{\mu^2 N_k}\bigg)^{1/3} 
\label{energy_ratio}
\end{align}  
where, 
\begin{align}
f(x) = \frac 1 3\bigg[x - \bigg(\frac{2}{\Delta}\bigg)^{1/3}x(x -6) - \bigg(\frac{\Delta}{2}\bigg)^{1/3} \bigg] 
\end{align}
and 
\begin{align}
\Delta = -27x +18x^2 -2x^3 +3\sqrt{3}\sqrt{27x^2 -4x^3} .        
\end{align}
Here we have defined $x=3\mu^2$ and $\mu = \sqrt{\beta/\alpha}$. Finally, using eqn.\,\eqref{vnpot2}, eqn.\,\eqref{hubb_brane} and eqn.\,\eqref{energy_ratio} in eqn.~\eqref{reh_eq} we can write $T_{\text{reh}}$  as 
\begin{align}
T_{\text{reh}}= \text{e}^{3(N_k - 67.35)}~\bigg(\frac{k}{a_0 H_0}\bigg)^3~\frac 7 6 f(x) ~ \bigg[\frac{1}{(3\alpha)^{-1/3}\mu^2\pi^2 M_{5}^3 A_s}\bigg]\times\frac{3}{r} , \label{trehn}
\end{align} \\
where we considered $g_{\star,\text{reh}}=g_{\text{s,reh}}$ \emph{i.e.} assuming no entropy production after the reheating epoch. Here the tensor-to-scalar ratio for the potential of eqn.\,\eqref{vnpot2} can be written as \cite{Herrera:2019xhs}
\begin{align}
r = \frac{48\tau\alpha^{1/3}}{\big(3N_{k}^2(1+\mu^2 N_k)\big)^{2/3}}\label{attr_r}
\end{align}
 Therefore, we have finally arrived at the expression for the reheating temperature in terms of the potential parameters as well as the brane tension $\tau$ (or $M_5$). Now we can study the variation of $T_{\text{reh}}$ with respect to any one of the inflationary observables.
 
In Fig.~\ref{reha_swamp} we plot the reheating temperature as a function of tensor-to-scalar ratio $(r)$ for different values of the  parameter $\mu$ and number of e-foldings $N_k$. From this plot, we see that for fixed values of $N_k$, the reheating temperature has a mild dependence on the parameter $\mu$. In fact, $T_{\text{reh}}$ increases very slightly as we increase $\mu$ from $0.2$ (right) to $1.5$ (left). This feature is also reflected in Fig.~\ref{ns_variation} where we have plotted $T_{\text{reh}}$ as a function of another inflationary observable --- the scalar spectral index $(n_s)$. In particular, we found that for $\mu\gtrsim 1.5$, the reheating temperature becomes imaginary. Therefore, for a meaningful physical interpretation of $T_{\text{reh}}$, we must restrict $\mu$ somewhere between $0.2 \lesssim \mu \lesssim 1.5$ as shown in fig.~\ref{reha_swamp}. 
Here the lower limit comes from the requirement of the potential of eqn.~\eqref{vnpot2} to be an attractor type, \emph{i.e.} $\alpha/\beta < N_k$ or $\mu \gtrsim 0.19 $ (for $N_k$ in the range $45 - 60$).
Also, in~\cite{Herrera:2019xhs} it was shown that for this potential, the parameter $\alpha$ depends on $N_k, \tau$ and scalar curvature perturbation $A_s$. Then using $A_s \sim 10^{-9}$ from observation and taking $N_k$ within the regime $45\lesssim N_k\lesssim 60$, one gets $2\times 10^9(\kappa/\tau)^2 \lesssim \alpha \lesssim 3\times 10^9 (\kappa/\tau)^2$. In Figs.~\ref{reha_swamp} and ~\ref{ns_variation} we have used this regime of $N_k$ and have considerd $\alpha\sim\mathcal{O}(10^9)(\kappa/\tau)^2$. Moreover, we have taken $\tau\sim 10^{-12}M_{\text{pl}}^4$ which means $M_5\sim 10^{-2}M_{\text{pl}}$. This choice of $\tau$ agrees with the upper limit on $M_5$ obtained in the last section. Later we have shown it is also consistent with the bound on $M_5$ obtained from the consideration of critical $T_{\text{reh}}$ for this model.




Now due to energy conservation, energy density at the end of reheating, $\rho_{\text{reh}}$ must to be smaller than the energy density at the end of inflation $\rho_{\text{end}}$. In other words, this means $\rho_{\text{reh}}\le \big(\frac{\rho_{\text{end}}}{\rho_{\text{inf}}}\big)\rho_{\text{inf}}$, \emph{i.e.} 
$\frac{\pi^2}{30} g_{\text{reh}} T_{\text{reh}}^{~4} \le \big(\frac{\rho_{\text{end}}}{\rho_{\text{inf}}}\big)\rho_{\text{inf}}$. Here the equality sign holds for the maximum possible value of the reheating temperature.
But at high energy regime, $\rho_{\text{inf}}$ is related to the Hubble parameter as $H_{k}^2 \approx \frac{\rho_{\text{inf}}^2}{6\tau M_{\text{pl}}^2}$. Therefore, after a little simplification we finally obtain, 
\begin{align}
T_{\text{reh}}^{\text{cr}} = \bigg(\frac{30}{\pi^2 g_{\text{reh}}}\bigg)^{1/4} \bigg(\frac{\rho_{\text{end}}}{\rho_{\text{inf}}}\bigg)^{1/4}
(6\tau)^{1/8}(H_k M_{\text{pl}})^{1/4}
\end{align}
Here $T_{\text{reh}}^{\text{cr}}$ is the upper bound on the rehating temperature, known as the critical reheating temperature.
Plugging in $\rho_{\text{end}}/\rho_{\text{inf}}$ from eqn.~\eqref{rho_ra}, $H_k$ from eqn.~\eqref{hubb_brane} and taking $g_{\text{reh}} =106.75$, the critical value for the reheating temperature turns out to  $T_{\text{reh}}^{\text{cr}} \simeq (1.31-2.49)\times 10^{15}$ GeV, which is indicated by the dotted purple line in fig.~\ref{reha_swamp}. 
Here, the small variation in the critical value of $T_{\text{reh}}$ is due to its dependence on $\mu$ for fixed $M_5$, whereas $T_{\text{reh}}^{\text{cr}} $ is found to be almost independent of $N_k$. Moreover, for the entire range of values of $\mu$, the upper bound on $N_k$ is such that $N_k\lesssim 59.7$.

\begin{figure}[t]
    \centering
    \includegraphics[width=8.1cm]{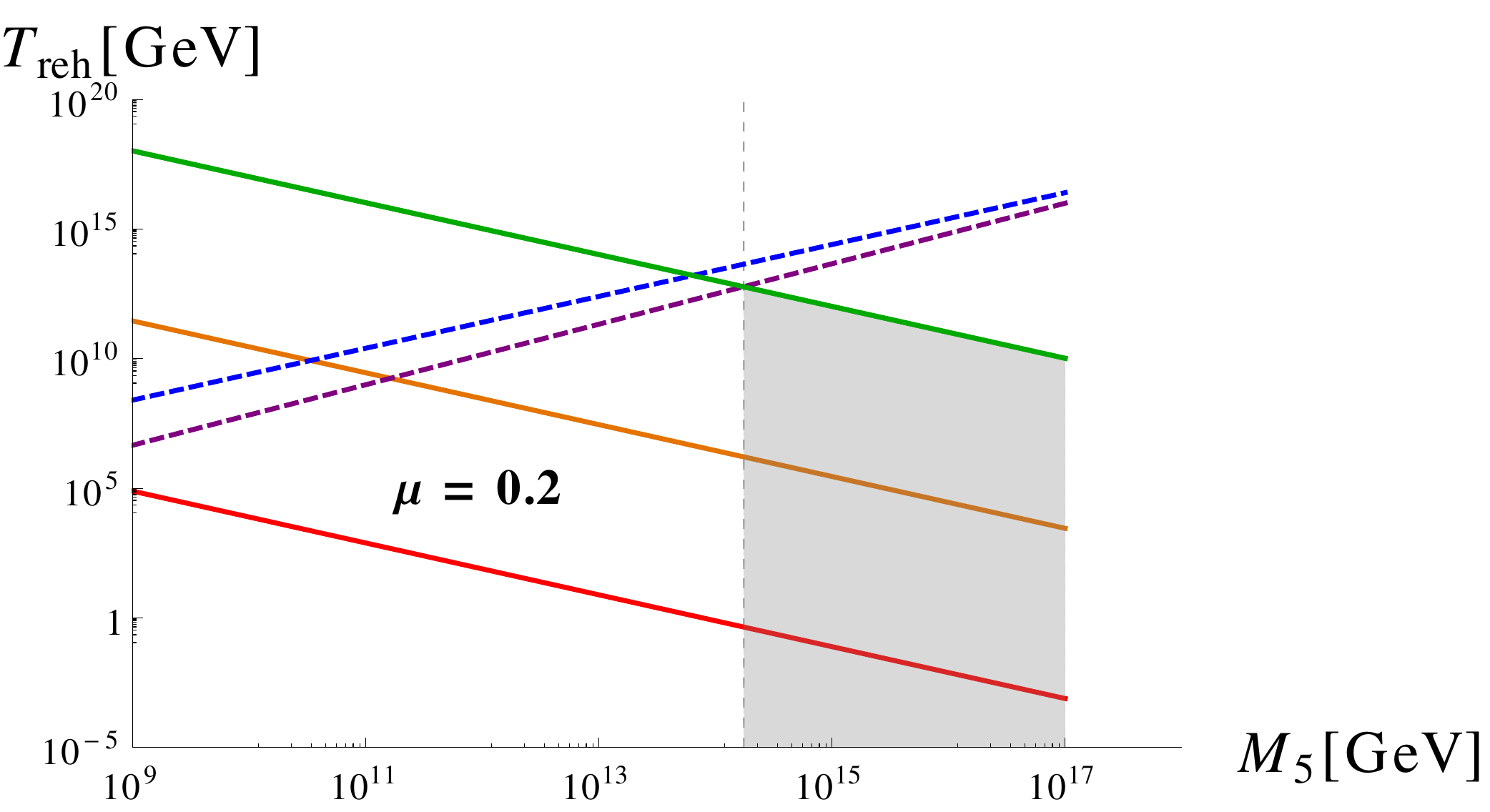} 
	\includegraphics[width=8.1cm]{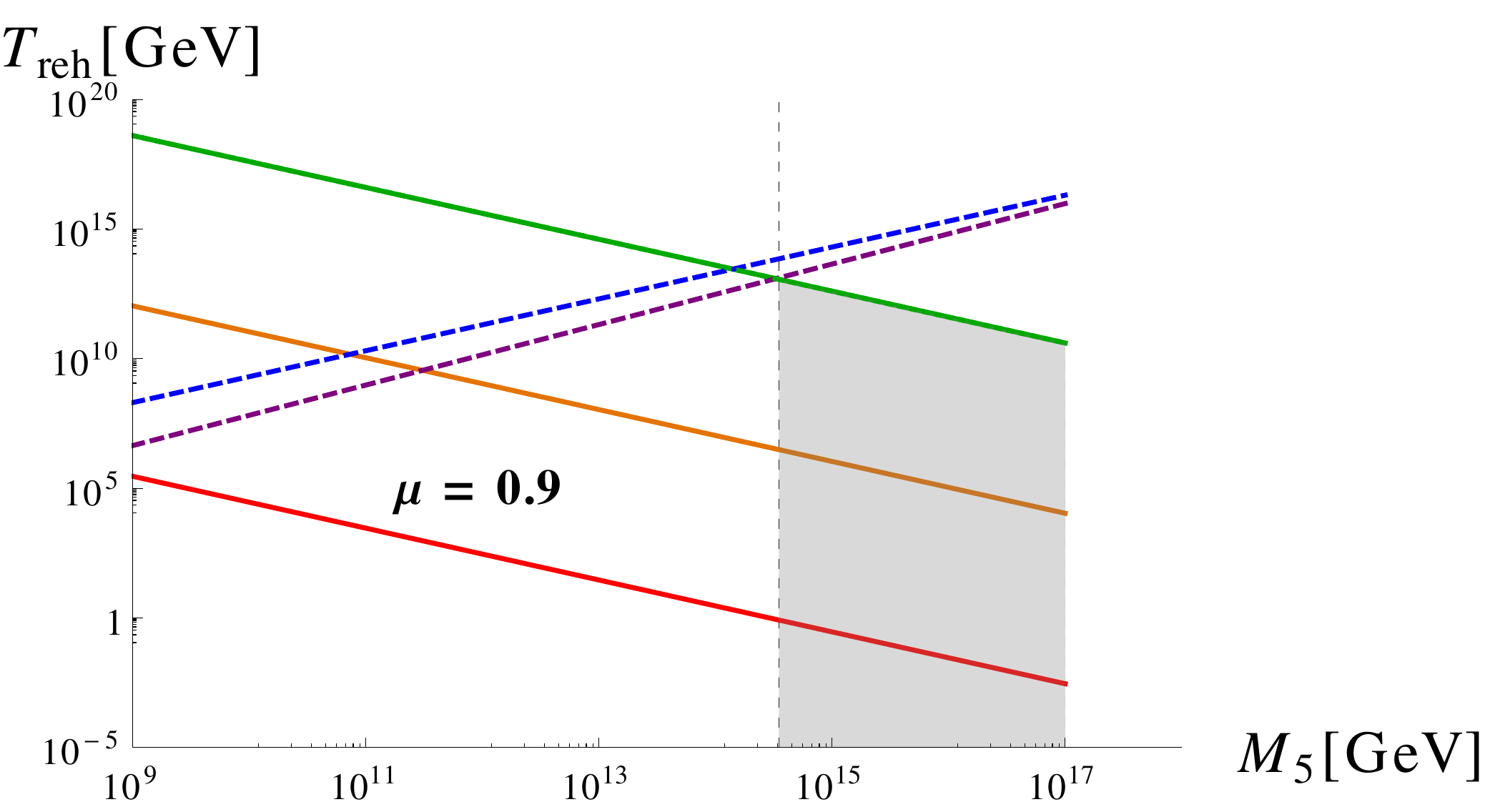} ~ 

\vspace{0.5cm}

	\includegraphics[width=8.1cm]{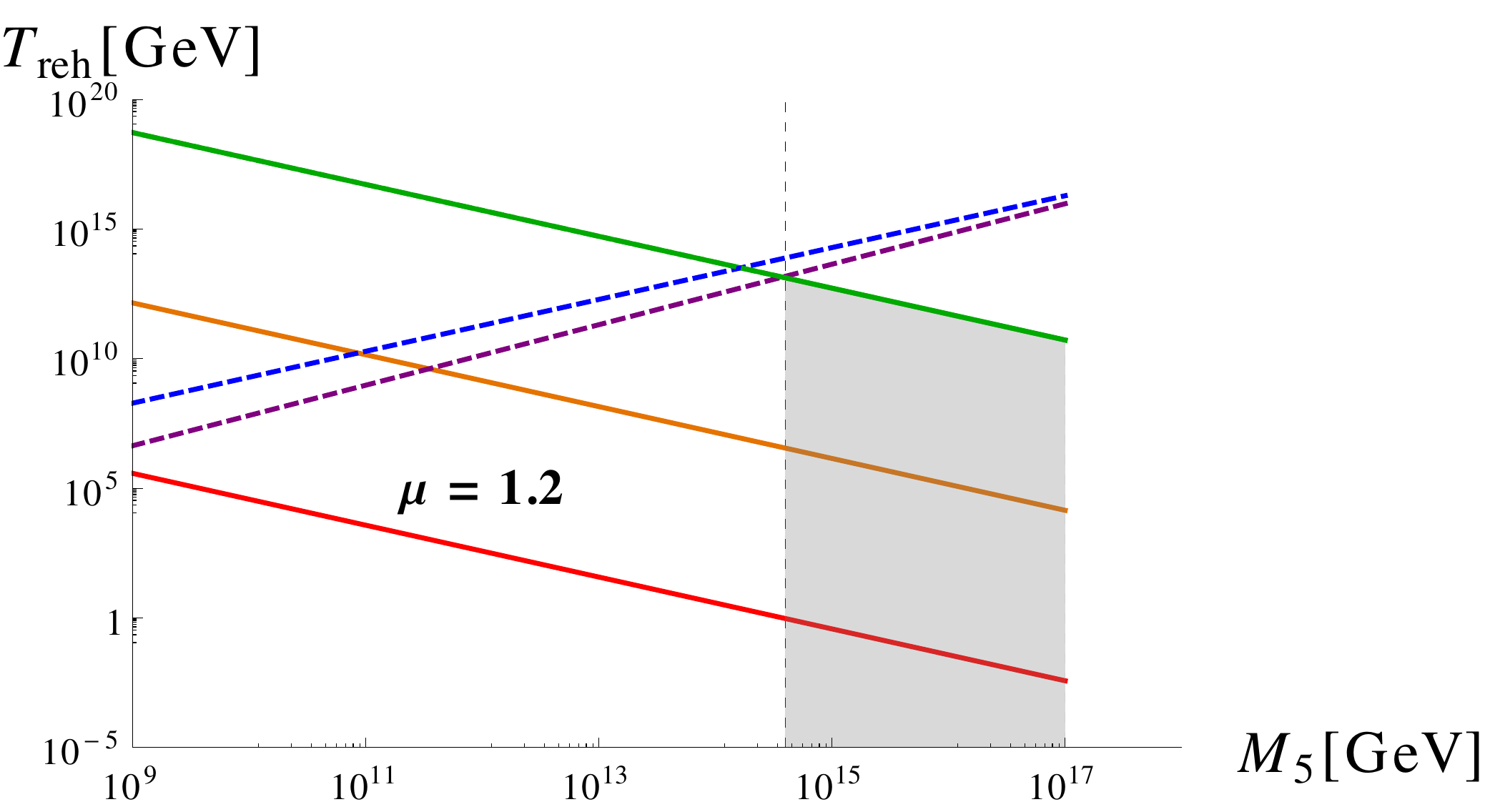} 
              \caption{\small Plot of $T_{\text{reh}}$ (at $k=0.002$ Mpc$^{-1}$) as a function $M_5$ for three different choices of the parameter $\mu$. In each of these plots, the colored lines \emph{i.e.} green, orange and red are showing the variation of $T_{\text{reh}}$ with respect to $M_5$ corresponding to $N_k = 45,50$, and $55$ respectively. All axes are in logarithmic scale. 
The critical reheat temperature (dashed purple) and the inflationary potential $V_{\text{inf}}^{1/4}$ (dashed blue) are plotted in each case to ensure that the inequality $T_{\text{reh}}\lesssim V_{\text{inf}}^{1/4}$ is satisfied everywhere. The shaded gray regime indicates that, for $0.2\lesssim\mu\lesssim 1.5$, the allowed range of $M_5$ lies between $10^{14}\lesssim M_5 \lesssim 10^{17}$ in GeV unit.  } 
\label{mu_variation}
\end{figure} 


In this braneworld reconstructed potential, both $T_{\text{reh}}$ and it's critical value are functions of the brane tension $\tau$ (through $M_5$) as well as the model parameter $\mu$. Fig.~\ref{mu_variation} shows the joint variation of $T_{\text{reh}}$ and $T_{\text{reh}}^{\text{cr}}$ with respect to $M_5$ for three different values of the model parameter $\mu$. For a given value of $\mu$, $T_{\text{reh}}$ is plotted with respect to $M_5$ for $N_k = 55$ (green), 50 (orange) and 45 (red). The critical reheat temperature $T_{\text{reh}}^{\text{cr}}$ and inflationary energy density $V_{\text{inf}}^{1/4}$ are also shown simultaneously to reflect the fact that the energy hierarchy $T_{\text{reh}} < V_{\text{inf}}^{1/4}$ is satisfied in the entire parameter regime. From the pont of intersection of $T_{\text{reh}}$ and $T_{\text{reh}}^{\text{cr}}$, we get a bound on the 5-D Planck mass $M_5$ and it is $10^{14}\lesssim M_5 \lesssim 10^{17}$ in units of GeV (shown by the shaded gray region).



\section{Conclusion}
\label{sec:conl}
In this work we have dealt with the generalised reconstruction of the RS $\rm II$ braneworld inflationary potential from the flow equation approach. Though this approach was adopted to reconstruct inflationary models earlier in~\cite{Ramirez:2004fb}, here for the first time, we have used the information from the reconstructed potential and taken it forward to analyse the era of reheating. We found an upper bound on the reheating temperature and it is shown that this critical limit of the reheating temperature is quite stringent.

It will be interesting to explore, how much our result is sensitive to the equation of state parameter ($w_{re}$). It would be interesting to tackle the specialities such as gravitino overproduction problem in the braneworld \cite{khlopov1, khlopov2, khlopov3, copeland} in a future study.\\
From this bound on  the reheating temperature, one can also constrain the value of $M_5$ (5-D Planck mass). The bound $\mu\leq 1.5$, to obtain real value of reheating temperature, limits the largest possible $M_5<10^{-1}M_{\text{pl}}$ from Eq.~\ref{mfr}, which agrees very well with the upper bound on $M_5$ obtained from $T^{\rm cr}_{\rm reh}$ in the last section. The result obtained here can be compared with some seminal works where constraints on the brane tension is computed \cite{gergely1,gergley2,gleiser1}. In some recent works \cite{rocha1,rocha2}, including
observational bounds on gravitational waves at LIGO and eLISA, strongest bounds for the finite brane tension is also computed. Thus the results obtained here can also be compared with the constraints on brane tension studied at from different areas of physics~\cite{salumbides, felipe, belotelov}. Therefore, it will be an exciting avenue to explore and compare those results to infer how better we can limit $M_5$ from a more general perspective.

On that note, following the work \cite{gergely1,gergley2}, an alternative approach has been taken in \cite{r1,r2,r3} where, the brane tension is expressed in terms of the cosmological time instead of the temperature. This alternative but equivalent approach presented in \cite{r1,r2,r3} is interesting to explore in our case. In summary, in this paper, for a generalised reconstructed potential we probed the reheating era indirectly and constrained the 5-D Planck mass ($M_5$) which represents the new scale of the braneworld physics. Our results show that this scale has a lower bound of roughly $10^{14}{~\rm GeV}$. This is counter intuitive from the point of view of the origin of this model to solve the hierarchy between the forces. However, to address this issue, a detailed study of the reheating mechanism must be done in order to ensure if there is some lowering of this bound. We would also like to point out, the swampland criterion\cite{vafa, ArkaniHamed, Obied,Agrawal,Kinney} recently created some noise in the community. It is been shown in \cite{yogesh}, that swampland problem can be avoided in case of $RS$ braneworld and that makes the model more interesting to cultivate, keeping the swampland criterion in mind and do the analysis for a reconstructed potential.
 Thus, we hope to return to this aspect in future with a further detailed analysis, where the mechanism of reheating will be studied in depth to see how it affects the stated results.\\
 \\
\textbf{Acknowledgements:}
SB is supported by institute post-doctoral fellowship from Physical Research Laboratory. SB is thankful to the Centre for Theoretical Physics, Jamia Millia Islamia, for their hospitality at later stages of this project. KD is supported by institute post-doctoral fellowship from S. N. Bose National Centre For basic Sciences, Kolkata. Work of MRG is supported by Department of Science and Technology, Government of India under the Grant Agreement number IF18-PH-228 (INSPIRE Faculty Award).  KD, MRG and SB acknowledge the Theory division, Saha Institute of Nuclear Physics, India, where this project was initiated.
\\

\end{document}